\documentclass[conference]{IEEEtran}

\usepackage{cite}
\usepackage{amsmath,amssymb,amsfonts}
\usepackage{algorithmic}
\usepackage{chronology}
\usepackage{graphicx}
\usepackage{textcomp}
\usepackage{xcolor}
\usepackage{makecell}
\usepackage{hyperref}
\usepackage{caption}
\usepackage{subcaption}
\usepackage{float}
\usepackage{url}

\def\BibTeX{{\rm B\kern-.05em{\sc i\kern-.025em b}\kern-.08em
    T\kern-.1667em\lower.7ex\hbox{E}\kern-.125emX}}
\begin{document}

\title{
Failures of public key infrastructure: 53 year survey 
}

\author{
\IEEEauthorblockN{Adrian-Tudor Dumitrescu}
\IEEEauthorblockA{ 
\textit{Delft University of Technology}\\
Delft, The Netherlands \\
A.T.Dumitrescu@student.tudelft.nl}
\and
\IEEEauthorblockN{Johan Pouwelse \small (thesis supervisor)}
\IEEEauthorblockA{
\textit{Delft University of Technology}\\
Delft, The Netherlands \\
J.A.Pouwelse@tudelft.nl}
}

\maketitle

\begin{abstract}
The Public Key Infrastructure existed in critical infrastructure systems since the expansion of the World Wide Web, but to this day its limitations have not been completely solved. With the rise of government-driven digital identity in Europe, it is more important than ever to understand how PKI can be an efficient frame for eID and to learn from mistakes encountered by other countries in such critical systems. This survey aims to analyze the literature on the problems and risks that PKI exhibits, establish a brief timeline of its evolution in the last decades and study how it was implemented in digital identity projects.
\end{abstract}

\section{Introduction}  \label{intro}
Digital identity is a rapidly growing field, driven by the increasing need for secure and trustworthy online transactions, prompting even governments to take action towards the future of the population. This transition reflects the profound impact of technology on how individuals perceive and manage their identities in an increasingly interconnected and online world. While the adoption of digital identity has yielded mixed outcomes, it bears the potential to endow individuals with social and economic empowerment, with the capacity to unlock economic value estimated to range between 3 and 13 percent of GDP by the year 2030\cite{bible}.

Digital ID systems, despite being promoted for development purposes, pose serious human rights risks and often suffer from implementation failures. These risks are acknowledged even by proponents of such systems. Unfortunately, there is a lack of comprehensive evidence and monitoring of their human rights impacts. Activists, journalists, and researchers have played a crucial role in documenting these impacts, particularly in cases like Aadhaar in India. The evidence gathered so far reveals that digital ID systems can result in various urgent human rights issues, including violations of the right to nationality, restrictions on access to healthcare, food, and social security, and a range of other concerns\cite{read}.

Public key exchange cryptography, a pivotal technological advancement articulated even more than 40 years ago\cite{diffie2022new}, underpins the security of public networks, enabling global communication and commerce. To establish trust and identity in digital communication, public keys, and implicitly private keys, must be associated with specific identities. This necessity led to the development of Public Key Infrastructures (PKI), which facilitate the issuance and storage of digital certificates.
These certificates verify that a public key corresponds to a particular entity. PKI offers a secure foundation for digital communication by providing authentication, encryption, and digital signatures through the management of cryptographic keys and certificates. It ensures the integrity of data, facilitates non-repudiation, and establishes trust in online transactions. Certificate authorities (CAs), trusted third parties, publish these certificates, connecting public keys to users via a private key. Public key cryptography has played a crucial role in establishing online identity, from traditional PKI and CAs to experiments like PGP's web of trust, and more recently, the blockchain ecosystem\cite{brief} that needs to authenticate the nodes of the networks and use different PKI approaches such as Multi-Layered Approach, Instant Karma PKI or Guardtime Approach \cite{pal2021key}. However, this relationship has its disadvantages in such that the shortcomings PKI brings can affect future digital ID infrastructures.

The interest of the European Union regarding the usage of digital ID has increased in recent periods, incorporating this vision in the EU developments and since 2021 drafting recommendations towards "a common Union Toolbox for a coordinated approach towards a European Digital Identity Framework"\cite{eueid}.
As Europe advances toward seamless digital verification, caution must be taken not to create a surveillance state and a centralized 'digital identity' as it has the potential to erroneously label legitimate users as 'bad actors'. Accumulating sensitive digital information raises security concerns, and misidentification risks hindering legitimate users. Digital verification, like secure blockchain, offers advantages over paper documentation, reducing forgery and theft risks. To succeed, these digital systems must comply with the GDPR and align with the European Commission’s 2020 data strategy, promoting secure and universally usable digital identities within common European data spaces as also stated by William Echikson in "Europe’s Digital Identification Opportunity"\cite{echikson2020europe}. However, these risks and potential problems ought not to stop the evolution of digital identity that is currently occurring in the world. With the recent pandemic and migration crisis Europe is confronting, adopting a unified electronic identification can help with a potential reduction in customer onboarding costs of 90\% \cite{bible}. In the end, rather than dividing nations, citizens can prove their identity and "share electronic documents from their European Digital Identity wallets. They will be able to access online services with their national digital identification, which will be recognized throughout Europe"\cite{ivic2022digital}.

This survey attempts to explore and reason the problems the PKI systems had and still exhibit after a long time from its introduction, alongside a brief history of evolution in view, ending with electronic ID implementations and failures from different countries. In section \ref{pki problems} we discuss the big problems of PKI and what risk it presents in incorporating it in different domains. Next,
in section \ref{alternatives}, we attempt to define a timeline of possible infrastructure alternatives that try to solve in part the PKI shortcomings presented and present different views of the architecture. In section IV, we discuss how different countries in the world tried to implement digital identity and sometimes failed.
PKI history starts with research report No. 3006 which presented the \emph{possibility of secure non-secret encryption}. This report was written in Jan 1970 and classified as \textbf{secret} within the CESG British government laboratory\cite{ellis1970possibility}. Our survey shows the surprising difficulty of PKI realisation at scale. The European Commission is currently aiming for the largest PKI attempt in history as part of their \emph{digital decade} (€165 billion in funding\cite{EU_invest}). 
Historical evidence going back 53 years indicates the EU should proceed with caution.

\section{PKI problems and risks} \label{pki problems}
The fundamental component of Public Key Infrastructures, involving key exchange through the RSA cryptosystem, has faced various attacks since its inception. Achieving secure implementation of RSA is a challenging endeavor, underscoring the complexity associated with deploying public key cryptography.\cite{boneh1999twenty}
In recent years, PKIs have gained attention, with many organizations announcing their intention to provide certification services to the public. While some have successfully implemented PKIs, challenges leading to failures can be attributed to a variety of factors, including technical, economic, legal, and social considerations.\cite{why_failed}.
\begin{itemize}
\item \textbf{Technical Reasons}:

The technical landscape of PKIs is beset with complexities. Central to PKIs are public key (X.509) certificates, intricate and non-intuitive data structures. Their complexity poses substantial obstacles to deploying PKIs on a large scale, which is at odds with the direction of creating national or global digital identities. Furthermore, managing certificates, including tasks like key pair generation and certificate revocation, proves to be a daunting and error-prone undertaking. PKIs rely on globally unique X.500 Distinguished Names (DNs), which are often challenging to define and maintain resulting in death-by-complexity of its usage. Alternative models like SPKI and SDSI have struggled to gain widespread adoption. Additionally, cross-certification, the mutual recognition of Certificate Authorities (CAs), faces challenges due to variations in certification practices and a lack of incentives for cross-certification.

\item \textbf{Economical Reasons}:

Establishing and operating a PKI necessitates substantial investments in secure facilities, hardware, and personnel. Calculating the Return on Investment (ROI) for PKIs is intricate since they provide infrastructure rather than specific chargeable services. This intricacy makes building a sustainable business case for Certification Service Providers (CSPs) offering certificates a formidable task, given the high costs and limited revenue streams.

\item \textbf{Legal Reasons}:

PKIs raise questions about liability, with certificate providers potentially held accountable for damages resulting from misuse or technical failures. As further elaborated in the subsequent discussion of risks, the inability to repudiate digitally signed statements can lead to predicaments for certificate owners who may be unjustly held responsible for actions they did not authorize.

\item \textbf{Social Reasons}:

Certificates are sometimes misunderstood as a means to establish trust, but trust in digital relationships differs from real-world trust based on personal experiences with the level of trust we get from certificates often being overestimated. In addition, users often lack awareness of the vulnerabilities and risks associated with public key cryptography, accepting certificates without considering potential security implications.
\end{itemize}

As highlighted by Carl Ellison and Bruce Schneier in various risks associated with Public Key Infrastructure and the use of digital certificates, PKI is not a silver bullet for security and has potential pitfalls and challenges in its implementation\cite{10risks}. These risks are presented as:
\begin{itemize}
    \item \textbf{Trust in Certificates} \\
    The risk of misplaced trust in certificates issued by Certificate Authorities (CAs). Just because a CA is "trusted" doesn't mean you can necessarily trust a certificate for a specific purpose.

    \item \textbf{Identity Verification} \\
    Challenges in verifying the true identity of the certificate holder, particularly when relying on names or other identifiers.

    \item \textbf{Non-Repudiation} \\
    Legal issues surrounding non-repudiation, where individuals may be held legally responsible for actions taken with their private keys, even if those actions were not their own.

    \item \textbf{Security of Verifying Computers} \\ The need to ensure the security of computers used to verify certificates, as compromising these computers can lead to security risks.

    \item \textbf{Certificate Authority Authority} \\ Questions about the authority of CAs to grant specific authorizations in the certificates they issue.

    \item \textbf{User Involvement} \\
    The importance of considering users' understanding and actions when using certificates.

    \item \textbf{Registration Authorities} \\ 
    Risks associated with the use of Registration Authorities (RA) in addition to CAs in the certificate issuance process.

    \item \textbf{Certificate Holder Identification} \\
    Challenges in identifying the certificate holder, especially when relying on external sources like credit bureaus.

    \item \textbf{Certificate Practices} \\
    The importance of well-designed certificate practices and standards to ensure the proper use of certificates.

    \item \textbf{Single Sign-On} \\
    The need to consider how PKI integrates with other security practices, such as Single Sign-On (SSO), and the limitations of SSO in maintaining security.
\end{itemize}

\begin{figure*}[h]
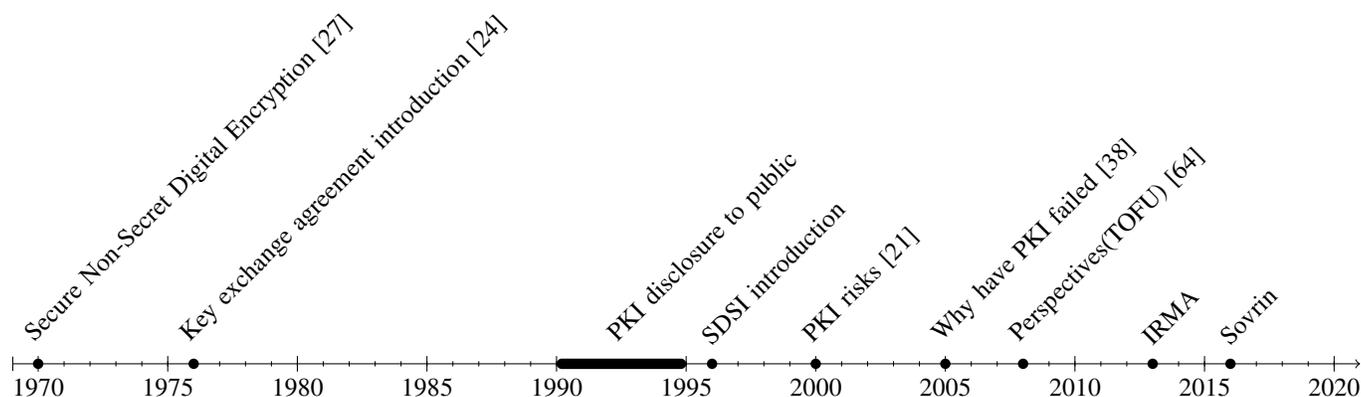

  \caption{Timeline of evolution and problem statements for PKI}
  \begin{chronology}[5]{1969}{2020}{\textwidth}
         \event{1970}{Secure Non-Secret Digital Encryption \cite{ellis1970possibility}}
         \event{1976}{Key exchange agreement introduction \cite{diffie2022new}}
         \event[1990]{1995}{PKI disclosure to public}
         \event{1996}{SDSI introduction}
         \event{2000}{PKI risks \cite{10risks}}
         \event{2005}{Why have PKI failed \cite{why_failed}}
         \event{2008}{Perspectives(TOFU)\cite{wendlandt2008perspectives}}
         \event{2013}{IRMA}
         \event{2016}{Sovrin}
    \end{chronology}
    \label{timeline}
\end{figure*}

\section{Evolution of PKI} \label{alternatives}
In this chapter, there are presented different views of how the Public Key Infrastructure can evolve and a brief history of alternative systems tried. PKI provides authentication, encryption, and digital signatures, ensuring secure communication, data integrity, and trust in
online transactions. From its introduction to the public, alongside the explosion of the World Wide Web, it was only considered from a centralized point of view that had the trust in the Certificate Authorities. The majority of the solutions presented are developing towards a decentralized view of the system, starting as early as 1996 with SDSI, that try to solve some problems from Chapter \ref{pki problems}. After less than 10 years from its public use, PKI shortcomings were addressed in KeyNote\cite{blaze1999keynote}, a trust-management system that uses a decentralized approach to handling public key infrastructure, allowing entities to manage their own keys and certificates addressing the scalability and flexibility issues associated with centralized PKIs. The full timeline of events can be observed in Figure \ref{timeline}.
\subsection{SDSI}
SDSI, or A Simple Distributed Security Infrastructure, is an innovative framework designed to address the complexities of security in distributed computing environments\cite{SDSI} tackling the first PKI problem presented while offering a robust and flexible solution, a first step in direction of SPKI(Simple public key infrastructure).

One of the most important features of SDSI is its simplicity. It achieves this through clear and intuitive mechanisms that focus on fundamental aspects of security by defining and representing security principles, establishing naming and addressing conventions, and expressing security policies.

By providing a straightforward means to define who or what can access resources, SDSI simplifies the task of managing access control. This clarity extends to naming and addressing, enabling a seamless way to locate and identify entities within a distributed network. SDSI recognizes that trust is a foundational element in security, and its framework allows for the establishment of trust among various entities within the system. At the same time, SDSI is the first step towards a decentralized public key infrastructure, allowing entities to make local decisions about access control based on their local policies and knowledge, rather than relying on a central authority for all decisions

While SDSI offers a good step towards a less problematic infrastructure in distributed environments, it's important to acknowledge that it may not be suitable for all use cases with an accent on digital identity. As state also by the creators of SDSI, "We feel [...] identity certificates must typically in the end be examined by people, to see if the name and other attributes given are consistent with the attributes known to the human reader"\cite{SDSI}, the problem of trust is transferred to the issuer(country in our case).

Over the years after SDSI design, multiple attempts have been made to use it in more practical ways to be able to overcome the problem of economic reasons. For example, in 1997, one year after SDSI release, a C library was created \cite{fredette1997implementation} to pave the way for its usage in different scenarios alongside a Java implementation in 1998\cite{morcos1998java} followed by an implementation of a secure web client using SPKI/SDSI certificates
\cite{ImplSDSI} in 2000, to meet the growing importance of the World Wide Web and with a case study on the effect on a company.

\subsection{Perspectives(TOFU)}
Trust-on-first-use(TOFU) is a strategy where, during the first encounter with a server or system, the user accepts and stores its public key without explicit verification. Subsequent connections are allowed only if the presented key matches the stored key, being a pragmatic approach to bootstrapping trust in a key-based authentication system. From its introduction to Secure Shell protocol, TOFU was seen as an improvement to the PKI ecosystem from 2008 with Perspectives\cite{wendlandt2008perspectives}, with no certificate authority needed to verify the identity of server owners and grant them certificates. The validity of a service’s key is determined by its existence on the network over time. \textit{Perspectives} system helps mitigate man-in-the-middle attacks by providing users with a more reliable basis for trust in the presented server public key.

However, the infrastructure inherits problems from the TOFU protocol, assuming that the initial connection is secure and that the user can trust the initial public key received during the first connection. If an attacker can compromise this initial connection, they may be able to present a malicious key that Perspectives would then consider as legitimate. The Perspectives system relies on a network of notary servers distributed across the Internet, presenting a new decentralized solution of PKI. Implementing and maintaining a network of notary servers can be complex and the effectiveness of Perspectives relies on widespread adoption and a sufficient number of notaries to provide diverse perspectives. Achieving and sustaining this level of adoption can be challenging and can increase the already complex infrastrcture.

\subsection{IRMA}
IRMA stands for "I Reveal My Attributes" and is a project aimed at implementing attribute-based identity management that seeks to address issues related to attributes, their possibilities, and challenges\cite{alpar2013towards}.

The paper acknowledges the existence of cryptographic techniques for secure and privacy-friendly attribute-based authentication, noting that recent advancements in smart card technology have made it possible to deploy attributes in practical scenarios. The concept of attributes \cite{alpar2013credential} is used broadly to describe the properties of individuals. These attributes may range from anonymous attributes (non-identifying), such as gender or age, to identifying attributes, like bank account or social security numbers. The paper highlights that while the underlying technology ensures full unlinkability, attribute values may allow for linkability expanding the range of application scenarios.
It relies on the Idemix technology and uses
personal smart cards as carriers of credentials and attributes

The extensive use of attributes within IRMA leads to dependencies between attributes, where the issuance of one attribute may depend on the verification of another. These dependencies give rise to a tree structure for attributes and raise questions about what should be considered "root" attributes that do not depend on others. These considerations have implications for societal identity structures, including pseudonym accounts.

The paper suggests the involvement of an independent, non-profit foundation to manage the IRMA scheme, set policy, and oversee certificate management for access to the card. This foundation would play a crucial role in addressing sensitive issues related to attribute management and policy solving perhaps the economic and legal problem of monopoly of a company on a scheme.

More work has been added to the IRMA project, with an implementation for smartphones \cite{alpar2017irma} in 2017 to facilitate its usage by ordinary people (with an app and QR codes) but also for service providers using standardized JSON Web Tokens. In 2019, solutions were proposed to contribute to ensuring the confidentiality and integrity of IRMA credentials in various scenarios. "Backup and Recovery of IRMA
Credentials"\cite{derksen2019backup} emphasizes that a recovery solution for IRMA should be designed as a backup and restore mechanism. To enhance portability and user-friendliness, the backup should be encrypted in a way that allows storage in any location without imposing a specific storage location on the user.

In this, key management is a crucial aspect of the design. The primary solution involves using a mnemonic phrase that can be written down on paper, an approach that does not require technical expertise and is understandable to users. Additionally, parts of the key are managed by trustees or a trusted institution as a second authentication factor to enhance security.

\subsection{Soverin}
Security requirements for digital identity systems mirror those of traditional paper credentials, encompassing compatibility, unforgeability, scalability, low latency, and revocation capabilities. Digital identity systems offer advantages like minimal dependencies, privacy/anonymity, unlinkability, and selective disclosure, providing a level of control impossible in paper-based systems.

Privacy-oriented digital identity schemes, such as U-Prove and Idemix, have been proposed but face challenges in widespread adoption due to issues like compatibility and scalability. For that, Sovrin is a system that integrates anonymous credentials with revocation, emphasizing privacy, unforgeability, performance, and unlinkability. The implementation incorporates a distributed ledger inspired by Ethereum and Byzantine Fault Tolerant (BFT) protocols for scalability \cite{windley2016sovrin}.

Sovrin employs anonymous credentials based on zero-knowledge proofs, providing unlinkability and features like delegation and revocation. Privacy concerns are associated with revocation, but in \cite{khovratovich2017sovrin} paper, attribute-based sharding are proposed to enhance privacy during the revocation process(and closing the gap to IRMA). The revocation methods involve cryptographic accumulators for efficiency.

Overall, Sovrin aims to address privacy and security concerns in digital identity systems through its innovative design and implementation and states from its requirements "self-sovereign identity, where every person, organization, or thing can have its own truly independent digital identity that no other person,
company, or government can take away"\cite{reed2016technical}.
Furthermore, the paper explains what most distinguishes Sovrin as a distributed identity system: it is the first public permissioned ledger. The stack of the technology has 3 important levels: Sovrin Ledger, Sovrin Agents and Sovrin Clients.

In a comparison between IRMA and Sovrin\cite{nauta2019self}, adopting Sovrin is considered challenging for both service providers and credential users and, like the PKI, its commercial value can be overseen. At the same time, Sovrin is a complex project and still in progress with its documentation, being an open source, being somewhat scattered around. However, Sovrin has an advantage over IRMA in deployment in such that service providers do not need to host any server because of the Sovrin Ledger. Regarding the digital identity problem, Sovrin has been cited as a possible solution for the technology needed in such schemes\cite{windley2021sovrin}.

\section{National digital identity implementations} \label{countries}
In recent years, worldwide there have been multiple attempts to create a national digital identity for its citizens sometimes expanding to inter-state agreements and continental recognition such as the eID. The majority of these attempts incorporated the PKI and blockchain ecosystem and tried to strengthen security with interviews(Estonia) or biometric data. PKI in e-Gov is not a new idea, being present since 2000\cite{brands2000rethinking}, facilitating social security and economic growth. However, even the latest implementations acknowledge the shortcomings of the PKI. For example, the UAE digital ID project was started based on such an infrastructure and even after its release and years of research the vulnerabilities presented in section \ref{pki problems} still persist\cite{hableel2013public} with the focus being on a lack of business value, business requirements and business integration issues alongside "much confusion about the full scope of this
project"\cite{al2012pki}. The risk in such systems may also explain the decrease in rapid adoption of digital identity solutions that started 'promising' like the EU nordic common eID \cite{hansteen2016nordic} project started in 2015 with a set timeline that was not continued until 2023 yet. But these failures did not stop other countries from fully embracing forms of digital identity and the benefits of such adoptions have been studied more in recent years (\cite{third2018government}, \cite{sullivan2019blockchain}, \cite{bible}) to determine the financial impact. Switzerland represents a specific case where the eID project started in 2020 was discontinued because of a referendum in which citizens were more afraid of their private information in the digital world. Controlled by the state but provided mainly by private companies, the project is redesigned to be more user-centered(self-sovereign identity) and to be adopted from 2023 onward. Moreover, this showed that not all countries are ready to be eIDAS compliant within Europe, with Switzerland opting for the Swiss Federal Law on Electronic Signatures. Here we present some cases in the world of a national electronic ID.

\begin{table*}[ht]
\caption{Overview of the ranking countries eID.}
\centering
\begin{tabular}{ |c|c|c|c|c|c|} 
 \hline
 Year & Country & Technology & Managing entity & Mistakes made & Mistakes to avoid \\
 \hline
 2007 & Estonia & \makecell{KSI \\blockchain} & \makecell{Information System Authority \\ and SK ID Solutions} & \makecell{Third-party infrastructure with \\ hardware vulnerabilities; Patented \\ and no open-source}
 & \makecell{Predatory monopoly and closed source; \\ Rely on commercial entity for \\country critical infrastructure} \\ 
 \hline
 2010 & India & PKI/HSM & \makecell{UIDAI - autonomous \\ government agency} & \makecell{Single point of failure \\ Social security concerns \\ Low-security; Chipless card} & \makecell{Only biometric system not viable; \\ No centralized database; \\ Social discrimination} \\
 \hline
 2010 & Germany & PKI and RFID & \makecell{Federal Ministry of the \\ Interior and The Federal Office \\ for Information Security} & \makecell{Long public development with \\ multiple failed projects; \\ Loss of public trust} & \makecell{Non-Decisive and multiple large scale \\ tests; No incentive to use; \\ Compatibility issues} \\
 \hline
 2013 & Peru & PKI & \makecell{RENIEC - autonomous \\ constitutional body} & \makecell{Sell information to 3rd parties \\ Low rural adoption and information \\ Vulnerable database} & \makecell{Selling private information to companies; \\ Put a price on confidentiality \\ of citizen information} \\
 \hline
 2016 & Italy & \makecell{PKI, NFC \\ Single sign-on} & \makecell{Italian Ministry of the Interior \\ and AgID government agency} & \makecell{2 digital IDs in circulation \\ Not expanding usage to government \\ institutions; Only 1 CA} & \makecell{Not keeping multiple electronic IDs; \\ Monopoly and bottleneck \\ of 1 CA} \\
 \hline
   2020 &Switzerland& Discontinued & Project discontinued & \makecell{Not eIDAS compliant \\ Very harsh legal framework \\ No public awareness} & \makecell{Limit use to national not European; \\ Dissolution of project before test \\ Lack of trust from population} \\
 \hline
   2021 & UAE & \makecell{PKI \\ Single sign-on} & \makecell{Telecommunications Regulatory \\ Authority - federal agency} & \makecell{Fast mandatory adoption \\ in critical sectors \\ PKI scalability problem} & \makecell{Phone incompatibility \\ One-point of failure} \\ 
   \hline
 2022 & Canada & PKI & \makecell{Provincial Identity Information \\ Management(IDIM) \\ - British Columbia} & \makecell{Different account levels \\ to access different services \\ Regional use \\ Closed-source development} & \makecell{Development of regional IDs \\ No coordination \\ with national direction} \\ 
 \hline
\end{tabular}
\label{taxonomy}
\end{table*}

\subsection{India}
The Indian digital ID scheme, ‘Aadhaar’\cite{tyagi2020blockchain}, was first introduced in 2010 and has been linked to almost all states within the country\cite{banerjee2016aadhaar}. The project aims to provide a single, unique identifier that captures all the demographic and biometric details of every resident of India and is close to issuing Aadhaar digital ID cards at the same time as birth certificates. Even though the system tried to pass the risks and problems of the PKI, the biometric implementation of Aadhaar raised privacy concerns from the start and is regarded as a failure in the citizen-government relationship\cite{dixon2017failure}. This example shows that implementing a biometric ID scheme can be very delicate and comes with its risks as well that may balance the benefits, reported to be in 2018 at almost 10 million euros\cite{rao2019aadhaar}. In this type of ID scheme, privacy and integrity of the data is critical, a weakness that the indian system encountered in leakage of critical information\cite{sen2019decade}. Aadhaar is the world’s largest biometric identity database in the world, so it is vital that the privacy of individuals is not breached and the data is used only for the purpose for which it has been approved. Even after years of use the problem and skepticism still exists \cite{abhijeet2021decrypting}, with a big accent on availing welfare benefits, governmentalism, authentication without consent and dependency on connectivity. This proved that a pure biometric system is not viable for a national digital identity and relying on hardware security is fragile.

\subsection{Canada}
In contrast to the Aadhaar, Canada had a different strategy in adopting a general digital identity scheme. Rather like countries where there is a single centralized government agency that assumes the role of identity authority, Canada opted that no single federal government organization can provide digital identity for all persons within the jurisdiction but there are 14 different “roots of identity”\cite{abraham2020building} through which persons can establish who they are. In the paper "Building Trust: Lessons from Canada’s Approach to digital identity", the Canadian approach is described as being bold in terms of making friendly overtures to technological implementations of the latest development in solutions—self-sovereign identity, where there might not be a need for Web PKI but a more decentralized infrastructure. While self-sovereign identity has not been taken seriously by many other governments, two standards are being considered components: Verifiable Credentials and Decentralized Identifiers. The scheme for the national Canadian ID is seen as an improvement for the public and private sectors and has well-defined principles such as No Centralized Authority, Secured Blinded Infrastructure, Decentralized, Secured, and Private Data Architecture, Privacy and Controls and Book Keeping, Audit, and Billing\cite{canadian2018canada}. The development of a modern digital ID system is accelerated by the use case on the financial side, like open banking, with an estimation of a profit of 3 billion euros\cite{koeppl2020open} in its first year. The attempts to create such as system have already been made in the country with the Verified.me application by SecureKey Technologies Inc which is set to expand its use to multiple public/private institutions\cite{boysen2021decentralized}. This slow development determined certain states within Canada to implement their own digital identity, one example being the British Columbia Wallet\cite{watkins2007trust}, an idea whose requirements were formulated as early as 2007. The BCeID Wallet is described as "a single credential (username/password), issued in person at a government Point of Service location, to be used across a range of online services"\cite{lockton2009government}. This aimed to solve the stalemate situation and "a mess that the government did not want to experience"\cite{watkins2007trust}. However, this individual development of an identity system creates a gap between the national movement toward eID and regional developments. The BCeID can not be used outside the British Columbia state and in the latest version, the accounts have an associated type(Business, Personal, Basic) to be used depending on the institution and criticality of information provided.

\subsection{Germany}
Germany had one of the first\cite{noack2010introduction} officially declared eID schemes for eIDAS, following the User-centric model. It is based on the German national identity card and electronic residence permit. Due to the use of Extended Access Control (EAC), each SP requires an authorization certificate and either an own eID server or a corresponding eID service. In order to obtain such an authorization certificate, SPs usually need to apply first, including a substantial service fee. Public bodies are excluded from this rule, since every municipality is required to provide its services online by law. To make things more complicated, every federal state can have its own digital identity system, leading to a rather complex mostly SAML-based federation. For this, multiple projects have been proposed to experiment\cite{germID} like project OPTIMOS 2.0 which provides the ecosystem for the mobile eID, while the project Digital Identities tries to optimize the app. The mobile app AusweisApp2 can be used as long as the smartphone is equipped with near-field communication (NFC) capabilities and runs on either iOS or Android\cite{pohn2021eid}. In the end, after the experiments, German citizens are able to securely store their national ID on a SIM card in their smartphone and the mobile eID could be used to open a bank account, use eGovernment services and other online services. As such, the need for a card reader or a physical card to identify and authenticate citizens online was removed\cite{germen_eid}. However, one problem identified within the country was the very low usage of the eID in transactions and interactions even with a high adoption rate of the population\cite{Pthesis} some inhibitors being other identification and authentication methods, and involvement of the private sector. One important use will no longer be possible with the new ID card because the holder can no longer be forced to deposit the ID card or give up custody of it. With the new card entailing both the electronic proof of identity and a private cryptographic key for the generation of qualified electronic signatures, the sole ownership of it represents indispensable security. Thus, to prevent abuse, "it is no longer allowed to demand the ID card to be handed over at the front desk or gate of a building or used as a deposit when borrowing an object"\cite{hornung2010id}.

\subsection{Estonia}
The Estonian digital identity scheme is one of the only ones that distance itself from the original PKI infrastructure. It uses the Keyless Signatures Infrastructure (KSI) a globally distributed system for providing time-stamping and server-supported digital signature services that have a different architecture from PKI, incorporating an Aggregation Network, Core Cluster and Gateway\cite{buldas2013keyless}. Started as a project for electronic access to healthcare and residency systems, the case expanded to a full digital ID infrastructure, with the main reasons for implementing the Digital Signature Act and provide means for digital signing for Estonian residents\cite{martens2010electronic}. Since its introduction, the Estonian eID scheme has been praised (\cite{anthes2015estonia}, \cite{heller2017estonia}) for its adoption rate within both private and public sectors, but Estonia’s digital success is not about other "digital offerings such as digital democracy, citizen engagement, or digitally transforming public services such as the welfare state"\cite{kattel2019estonia} and disconnect between technological infrastructure and degree of digital penetration
alongside the small size of the population(and of the data) are often overlooked. Also from the policymaker perspective, there are identified challenges related to issues like implementation, (national) legislation, interpretation, compliance and communication. A crucial eIDAS implementation barrier is the lack of the EU common identifier and Estonia's scheme seems to further away even more\cite{lips2020eidas}. Estonia is one of the first countries that enabled E-Voting with the help of the digital ID\cite{goede2019estonia} with data	stored in a decentralized	fashion in over 360	databases in which all information	from local hosts is linked through a specific infrastructure, X-Road, that, however, presents a single point of failure for the whole eGov data transfer to stop. At the same time, Estonia decided to put a commercial entity in control of its critical infrastructure with few opportunities of change in the system. Moreover, the custom blockchain used is patented, without any open-source material, and can manifest hidden vulnerabilities like the crisis from 2017 \cite{parsovs2020solving} in which almost all cards (800.000) were affected.

\subsection{Peru}
The National Electronic ID Card (DNIe) of Peru, issued by the National Registry of Identification and Civil Status (RENIEC), was recognized as the top ID card in Latin America at the 2015 High Security Printing Latin American Conference held in Lima. RENIEC, functioning autonomously and responsible for civil registration, identification, and digital signatures, has distributed 30 million eIDs, covering nearly the entire population of the country. The DNIe grants Peruvian citizens a digital identity that can be verified both physically and virtually. It incorporates two digital certificates, enabling the cardholder to electronically sign documents with the same legal weight as a handwritten signature. Peru's eID adheres to the ISO/IEC-7816 standard, and its biometrics system aligns with ISO/IEC 19794\cite{grassi2017draft}. It implements the cryptographic methods and X.509 digital certificates defined by the Public Key Infrastructure (PKI) and comes with its known risks. First introduced in 2013, the specifications are being analyzed for a new form in terms of the card, hardware (subdivided into the antenna, chip, and memory), and software (subdivided into the operating system, applications, middleware, and complements \cite{quiroz2020requirements}. Similar to the Indian scheme, Peru's digital ID can be used for biometric identification but in this case is not a requirement and not used at large and even though the adoption rate for the population is almost 99\% there is still lack of a remote access of the e-ID \cite{gelb2018identification}. In the analysis, there are also presented possible risks for the future of the Peruvian digital car, such as making ID enrollment a prerequisite in areas with low coverage.

\subsection{Italy}
Italy presents an interesting case with the fact that it currently has 2 different digital identification systems online. SPID(Public Digital Identity System) is an identity management framework that was launched in 2016, that facilitates accessing public administration online services through once-only digital identity (user \& password) generated by private Identity Providers. This is an example of a classic centralized PKI system that comes with the problems and risks presented in previous chapters alongside being launched by Private Identity Providers. Even before its public introduction, the scheme used by SPID has been criticized for information leakage about customers of identity providers\cite{buccafurri2015enhancing} and not relying on the Italian public system\cite{buccafurri2016implementing}, creating a complex ecosystem. The CIE(Electronic Identity Card) is a system launched in late 2016 by the government and managed by the Italian Ministry of Interior, the institution that represents the only Certificate Authority in the infrastructure. It has Single Sign-On features and has been updated to support NFC in accordance with the eIDAS. Being launched after SPID, the CIE is less used in public institutions and this competition determined the government to start a merging process between the 2 systems.

\section{Conclusion} \label{conc}
No mass deployment exists, more than 53 years after the discovery of \emph{the possibility of public key encryption} \cite{ellis1970possibility}. PKI still presents problems and risks in complexity, legal framework, lack of investment and social awareness. These issues should not shadow the importance of the infrastructure providing authentication, encryption, and digital signatures, ensuring secure communication, data integrity, and trust in online transactions. During these decades, the infrastructure evolved with different perspectives trying to change its design from a centralized view to a decentralized one. Even so, improvements are still being developed and studied\cite{iab-web-pki-problems-05} to cover as many vulnerabilities as possible, especially in the context of using PKI in national digital identity systems. These programs can unlock as much as 6\% of GDP in certain countries\cite{bible}, with Europe trying to achieve full digital coverage by 2030. Unfortunately, developing such critical infrastructure is prone to mistakes in different maturity levels\cite{arora2008national}. We created an overview of eID projects in different parts of the world \ref{taxonomy} and listed mistakes to avoid in implementing digital identity in the future path of the EU.
 
\bibliographystyle{plain} 
\bibliography{survey} 

\end{document}